\documentclass[biblatex, english]{lni}
\usepackage{booktabs}
\usepackage{enumitem}
\usepackage{fontawesome5}
\usepackage{array} 
\usepackage{float}
\usepackage{draftwatermark}
\SetWatermarkText{Preprint}
\SetWatermarkScale{1}
\SetWatermarkColor[gray]{0.9}

\usepackage[]{blindtext}

\begin{document}
\title[Explorable Authoring Requirements]{Evaluating Authoring Tools with the Explorable Authoring Requirements}
\author[1]{Frederic Salmen}{salmen@cs.rwth-aachen.de}{0009-0007-5206-1146}
\author[1]{Ulrik Schroeder}{schroeder@cs.rwth-aachen.de}{0000-0002-5178-8497}
\affil[1]{RWTH Aachen University\\Learning Technologies\\Ahornstr. 55\\52074 Aachen\\Germany}
\maketitle

\begin{abstract}
Explorables with interactive, multimodal content, openly available on the web, are a promising medium for education. Yet authoring such explorables requires web development expertise, excluding most educators and students from the authoring and remixing process. Some tools are available to reduce this barrier of entry and others are in development, making a method to evaluate these new tools necessary. On the basis of the software quality model ISO 25010, empirical results, and domain modeling, we derive the Explorable Authoring Requirements (EAR) as a requirements catalogue explorable authoring tools should implement. We then outline a future research design to operationalize EAR. 
\end{abstract}


\section{Motivation} \label{motivation}
\begin{figure}
    \centering
    \includegraphics[width=0.85\linewidth]{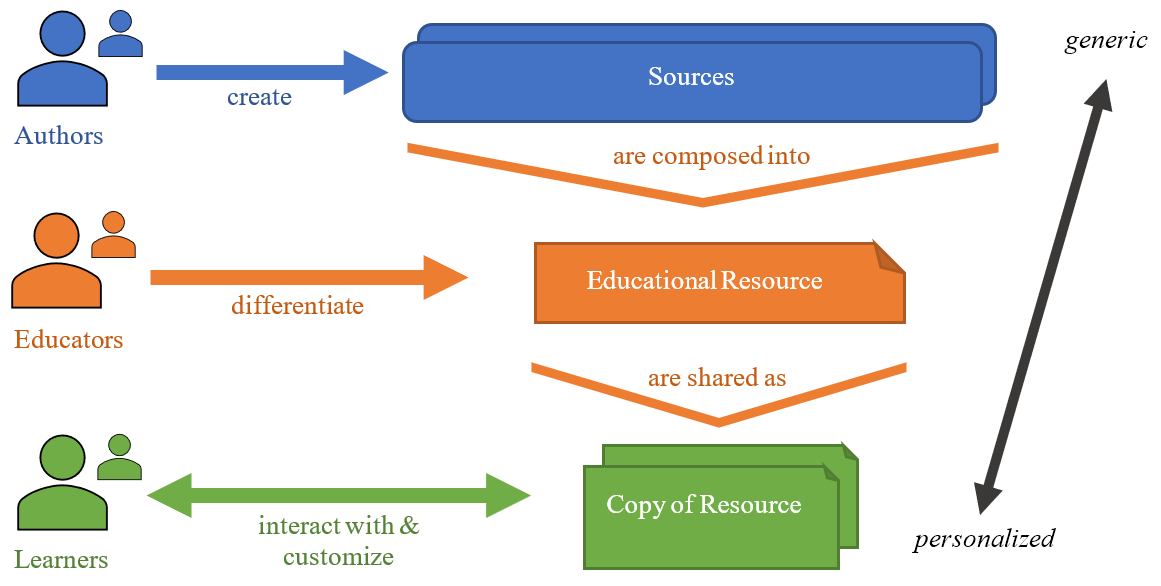}
    \caption{The remix workflow for authoring learning resources (reproduced from \cite{salmen_webwriter_2023})}
    \label{fig:remix}
\end{figure}

Consider an educator preparing a lesson. In addition to setting appropriate learning goals,  planning out learning activities for the students, and securing lesson results, the educator may need learning resources to use in the lesson. We define learning resources as material created for educational purposes.  The typical authoring process of such learning resources can be described with the \textit{remix workflow} (compare Fig.\ref{fig:remix}): Educators may use different sources created by original authors such as textbooks or websites and compose them into a resource, e.g. a worksheet. They may differentiate the resource considering their learners. Next, they are shared with the learners as copies, allowing learners to interact with and customize their copy, e.g. annotating the worksheet or completing tasks in their preferred order and tempo.

Educational technology research has produced a multitude of insights for learning resources: Simulations \cite{wieman_phet_2008}, learning videos \cite{brame_effective_2016}, serious games \cite{wouters_meta-analysis_2013}, etc. have proven beneficial for specific topics and situations. Research into multimedia learning \cite{mayer_multimedia_2009} shows that combining different media benefits learning. Active \& personalized learning \cite{shemshack_systematic_2020} proposes that the learner should take an active role in the learning process and that educators should differentiate learning resources to accommodate different learners. The Open Education Resources (OER) movement advocates for educators to become authors themselves and to share their resources with others \cite{muus-merholz_freie_2018}. \textit{Explorables} make use of the modern web platform to unite all of these benefits - as openly available web documents mixing text, images/audio/video, and interactive content \cite{victor_explorable_2011}. 

Still, interactivity can be seen as a double-edged sword. On one hand, highly interactive learning resources such as explorables can unite many benefits of different learning technologies. On the other hand, interactive content is hard to author, remix and edit, requiring time-intensive programming. Hohman et al. discuss this problem of explorables (called interactive articles in their terms), namely that "creating a successful interactive article is closer to building a website than writing a blog post" - requiring low-level web development knowledge incidental to the educational task at hand to succeed \cite{hohman_communicating_2020}. This disempowers both educators and students, who often lack the time and/or specialized knowledge to participate in the authoring process. That leaves the professional expertise of educators and the self-knowledge of students largely unused.

So how can we enable educators and students to participate in remixing interactive learning resources? One approach would be teaching them how to program, which may eliminate the `lack of technical expertise' problem. Another approach is to provide a specialized authoring tool focused on the remix workflow. These approaches are not exclusive, but complementary: Building technical competence in educators or teachers enables them to use tools in more advanced ways. This paper focuses on the second approach of an authoring tool.

There are different state-of-the-art authoring tools for creating explorables, most importantly H5P\footnote{https://h5p.org, accessed on 2024/02/29.} and Jupyter\footnote{https://jupyter.org, accessed on 2024/02/29.}. There are also recent research projects such as IdyllStudio \cite{conlen_idyll_2021} and WebWriter \cite{salmen_webwriter_2023}. In this paper, we want to introduce a way to evaluate such tools, allowing the research community to identify weaknesses in existing tools and to assess the impact of innovations in new tools. The main qualities we focus on are functionality and usability,  In summary, we pose the following research question:

\textbf{RQ1}: Which evaluable requirements can be found for functional, usable explorable authoring tools, considering educators and developers?

\section{Related Work} \label{relatedwork}
We first broaden our view from functionality and usability to consider if there are existing models to evaluate educational software. Early work by Jones et al. \cite{jones_contexts_1999} shows that there is significant overlap with evaluations in the HCI community, but that there are unique educational concerns that should not be marginalized by solely focusing on usability. Later work by Schleyer et al. highlights the hard problem of measuring learning outcomes, something educational technology shares with the whole field of research in education.  A bit more recently, some models specific to educational software were proposed \cite{jones_contexts_1999, escudeiro_evaluating_2009}, but they are still outdated, not suited to authoring tools, and are only limited extensions of more general software quality models.

As such, we opted to directly use general software quality models as a starting point, instead. These models enjoy extensive usage in software engineering research and practice \cite{wagner_software_2012} and are continually updated. The work by Wagner et al. further shows that quality models are typically adapted and extended with domain-specific details. Discounting quality models only suited to enterprise (e.g. quality gates or defect classes), the main models to consider are ISO 9126 and ISO 25010 \cite{international_standards_organization_isoiec_2011}. Both are international standards for software quality, and are also closely related, since ISO 25010 (part of the ISO 25000 set of norms) is the successor of ISO 9126. ISO 25010 divides software quality into eight characteristics, each with several sub-characteristics:
\begin{enumerate}
    \item Functional Suitability: Completeness, correctness, appropriateness
    \item Performance Efficiency: Time behavior, resource utilization, capacity
    \item Compatibility: Co-existence, interoperability
    \item Usability: Appropriateness recognisability, learnability, operability, user error protection, user interface aesthetics
    \item Reliability: Maturity, availability, fault tolerance, recoverability
    \item Security: Confidentiality, integrity, non-repudiation, accountability, authenticity
    \item Maintainability: Modularity, reusability, analyzability, modifiability, testability
    \item Portability: Adaptability, installability, replacability
\end{enumerate}

For the requirements of explorable authoring tools, there are only a few works to consider: Serth et al. \cite{serth_evaluating_2019} interviewed teachers, headmasters, and students, and found that interviewees saw the interactivity, possibility for personalization and the usability of existing solutions for digital worksheets in the subject of Computer Science to be lacking.  Salmen et al. \cite{salmen_webwriter_2023} ran a workshop-based study to derive a set of requirements and a first prototype for a GUI-based explorable authoring tool called WebWriter (see Tab. \ref{tab:explorable_requirements}).

\begin{table}[H]
    \centering
    \begin{tabular}{>{\raggedright\arraybackslash}p{0.3\linewidth}>{\raggedright\arraybackslash}p{0.6\linewidth}}
         \textbf{The system should...}& \textbf{Specific features from the workshop}\\
         …allow creating interactive multimedia usable for students& mix of content types, making interactive videos, \underline{embedding simulations}\\
         …be usable for teachers in remixing workflows& \underline{live preview}, \underline{undo/redo}, \underline{appealing visuals}, \underline{unlimited composition of content}, \underline{copy/paste external resources}\\
         …support personalized learning& specific content customization options, \underline{theming}, \underline{conditional branching}\\
         ...enable reuse \& retention& use on any operating system, import/export, usage within LMS, usage without LMS\\
    \end{tabular}
    \caption{Explorable authoring system requirements (\underline{Specifics not fulfilled by Lumi underlined}) (reproduced from \cite{salmen_webwriter_2023})}
    \label{tab:explorable_requirements}
\end{table}

Note that more discussion of related work follows in section \ref{ear} when the requirements model is described. 

\section{The Domain of Explorable Authoring} \label{domain}
In this section, we construct a basic model of what explorable authoring means. This will serve as a starting point on how to systematize requirements. First, we observe again that authoring and remixing explorables means authoring and remixing web documents.  The underlying data structure manipulated by an explorable authoring tool then is the Document Object Model (DOM). More details about web technologies can be found in introductory texts on web development (e.g. \cite{haverbeke_eloquent_2019}).

\begin{figure}[H]
    \centering
    \includegraphics[width=0.6\linewidth]{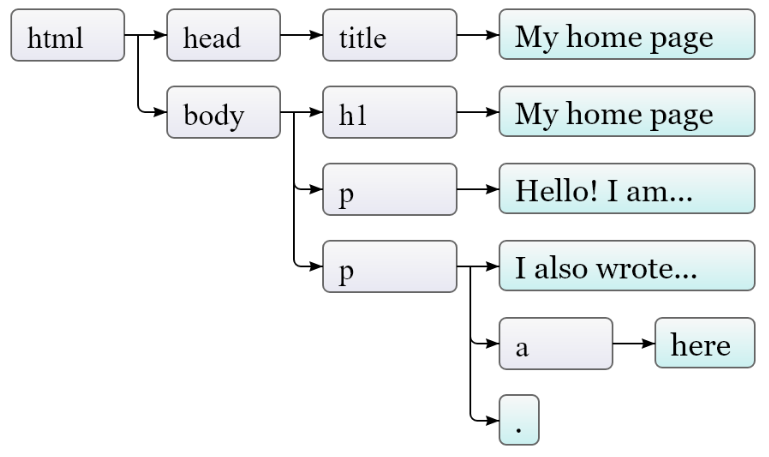}
    \caption{The document object model represented as a tree (reproduced from \href{https://eloquentjavascript.net/14_dom.html}{Chapter 14 of Eloquent JavaScript, 3rd Edition} by Marijn Haverbeke under \href{https://creativecommons.org/licenses/by-nc/4.0/deed}{CC-BY-NC})}
    \label{fig:dom-tree}
\end{figure}

\textbf{Basics}. The DOM represents any document as a tree (compare Fig. \ref{fig:dom-tree}) of \textit{nodes}. At the root is a document node with a single child node containing all other nodes, branching out into the head (for metadata) and body (for visible content). Nodes may either be elements which can have an ordered list of children and an unordered record of key-value pairs called attributes, or simple text nodes. There are some other node types not considered here since they are not important for the authoring process. Additionally, the way the DOM is rendered by the browser can be modified using style sheets (CSS), and the DOM can interact with itself or the browser using scripts (JS).

\textbf{Elements}. The web platform includes a large set of elements (110 at the time of writing), allowing authors to express many kinds of documents. At the same time, we can consider what content types state-of-the-art tools such as H5P, Jupyter, WebWriter, and IdyllStudio offer. In Tab. \ref{tab:content_types}, we try to categorize the available elements from an author's point of view to make this large set more easy to handle.
\begin{table}[H]
    \centering
    \begin{tabular}{ll}
         \textbf{Content Type}&  \textbf{Description}\\
 Metadata&Information about the document such as title, author, etc.\\
         Plain Text&  Text without any formatting\\
         Rich Text&  Bold, italic, underlined, strikethrough, link, etc.\\
         Headings&  Several levels of headings for to create a content hierarchy\\
 Sections&Semantic content blocks such as headers, footers, paragraphs, etc.\\
         Lists&  Ordered or unordered lists of items\\
         Tables&  Tabular data with rows and columns\\
 Forms&Interactive inputs such as text fields, checkboxes, etc.\\
 Disclosures&Toggleable elements such as dialogs\\
         Formulas&  Mathematical formulas (MathML)\\
 Graphics& Vector graphics (SVG) and graphics canvas\\
         Images&  Embedded image media\\
         Audio&  Embedded audio media\\
         Videos&  Embedded video media\\
 Documents&Embedded documents (web, PDF)\\
 Scripts&Code to execute on the document\\
 Themes&Styles to apply to the whole document\\
 Widgets& Custom interactive content (with JS/Web Components)\\
    \end{tabular}
    \caption{Content types of the web platform}
    \label{tab:content_types}
\end{table}
\textbf{Fragments}.  Fragments are effectively sub trees of the whole document tree, consisting of a segment of nodes. Fragments allow us to model editing operations as manipulations of aggregates of nodes, not just single nodes. Consider for example a fragment with multiple items of a list (e.g. "1. Eggs 2. Spam 3. Ham"), which is a complex sub tree (the list itself and the items each with text content). If we wanted to format the first and second item of the list as bold ("1. \textbf{Eggs} 2. \textbf{Spam} 3. Ham"), we'd need to select the two items (a fragment of the whole list and document) and wrap the text nodes of both items in 'bold' elements.

\textbf{Extensions}. The web platform also offers several ways of extension. Historically, a free-form use of scripts was the only practical way of extension. Scripts could be loaded into the DOM that modified existing elements to achieve custom behavior. A more modern, standardized way of extending the platform is available with Web Components, which encapsulate custom behavior and give access to many previously inaccessible parts of the web platform. The specific method of extension is not our concern here - in general, we mean an external module which changes the behavior and/or appearance of a document using an interface.

\section{Explorable Authoring Requirements (EAR)} \label{ear}
\begin{figure}
    \centering
    \includegraphics[width=1\linewidth]{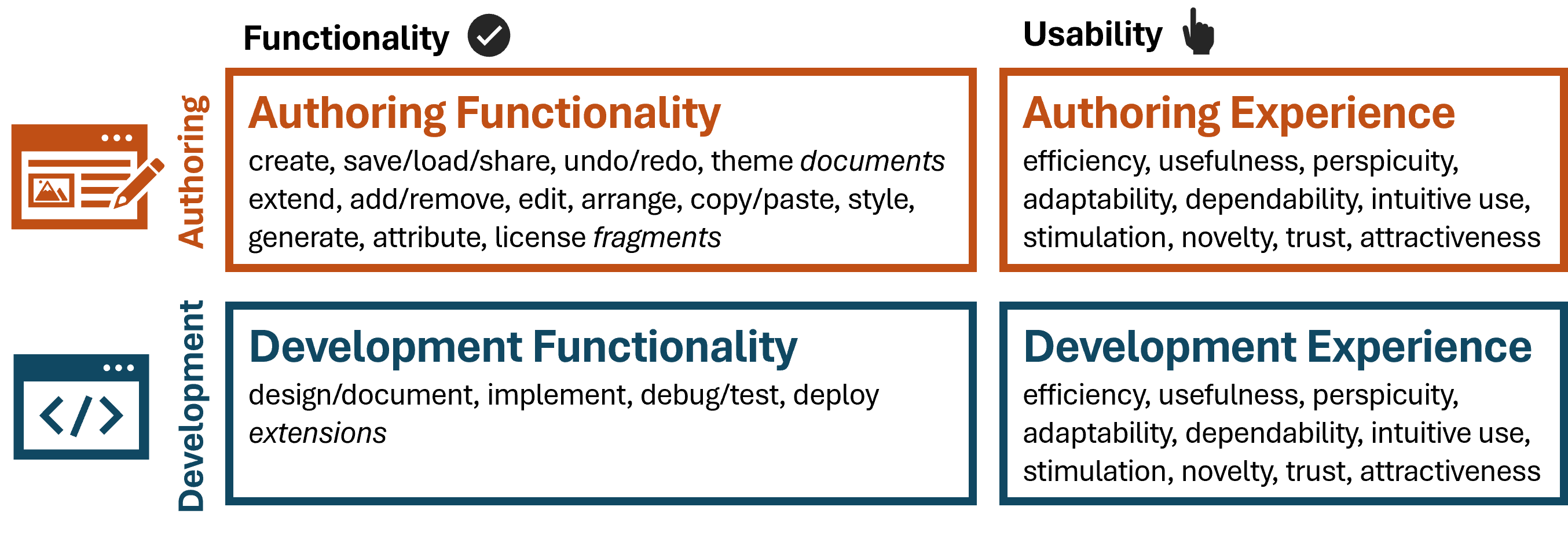}
    \caption{Explorable Authoring Requirements model, overview}
    \label{fig:ear-model}
\end{figure}

In this section, we will introduce a systematic catalogue of requirements for authoring explorables. This catalogue (compare Fig. \ref{fig:ear-model}) is a synthesis of the empirical results of Salmen et al. \cite{salmen_webwriter_2023} and Serth et al. \cite{serth_evaluating_2019}, the ISO 25010 characteristics, and the domain knowledge about explorable authoring. The catalogue has has two main perspectives: Authoring and development. Authoring refers to the editing of documents, and development refers to the editing of extensions. For each perspective, there is both the quality of functionality and the quality of usability to consider, resulting in the four requirements categories of \textit{authoring functionality}, \textit{authoring experience}, \textit{development functionality}, and \textit{development experience}.

\subsection{Authoring Functionality}
The authoring functionality category reflects the functional suitability characteristic of ISO 25010. It also makes use of the domain terms \textit{document} and \textit{fragment} to define a set of functionality intended to cover the requirements Salmen et al. and Serth et al. collected, as well as adding some "common sense" requirements for completeness:
\begin{enumerate}[leftmargin=2.25cm, label=EAR-AF\arabic*]
    \item \textit{Create}: Make new documents
    \item \textit{Save/Load/Share}: Store/restore document as/from external source
    \item \textit{Undo/Redo}: Restore a previous document state
    \item \textit{Extend}: Register new kinds of elements in the document
\item \textit{Add/remove}: Add fragments to the document
\item \textit{Edit}: Modify fragments
\item \textit{Arrange}: Reposition fragments in the document
\item \textit{Copy/paste}: Duplicate/add fragments
\item \textit{Style}: Change the appearance of a fragment
\item \textit{Generate}: Automatically generate a fragment from a prompt
\item \textit{Attribute}: Annotate a fragment with an attribution
    \item \textit{License}: Annotate a fragment with licensing terms
\end{enumerate}

The first requirements of the model are typical and straightforward for document-based work (AF1-AF4): Documents must be \textit{created} and \textit{saved/loaded} (in a file-based approach) or \textit{shared} (in a service-based approach). Another universal requirement is \textit{undo/redo}, As we established widgets as a content category, widgets also need to be made available for a document before they can be authored and remixed further, which we call \textit{extending} (AF4). 
The main part of the requirements concerns the manipulation of fragments (AF5-AF9). The complexity of an authoring tool lies in the multitude of element types (see Tab. \ref{tab:content_types}): While \textit{adding/removing}, \textit{editing}, \textit{arranging}, \textit{copying}, \textit{pasting} and \textit{styling} text is straightforward, this is significantly more difficult for trees of multimodal and/or interactive content. Simply put: Each function should work with every element type. To account for the emergence of generative AI models, we also introduce \textit{generate} as a functionality (AF10). For OER, we add \textit{attribute} and \textit{license} (AF11, AF12), since remixing content may legally or ethically require to attribute the original author, and in turn, content may need be licensed to enable reuse itself.

\subsection{Development Functionality}
In our previously introduced domain model of explorable authoring, we defined an extension as an external module which changes the behavior and/or appearance of a document using an interface. This makes creating an extension largely a process of software development, and similar requirements apply. Software engineering research and practice offers many different development processes tied to different methodologies, such as the waterfall model or agile approaches such as SCRUM \cite{sommerville_software_2011}. Because of the multitude of approaches to software engineering used today, authoring tools and their extension interfaces should be agnostic as to which one is used. Instead of prescribing specific approaches, authoring tools should focus on providing technical support for a common denominator of core functionality. 

The Rational Unified Process (RUP) model defines a set of processes (\cite{sommerville_software_2011}, Figure 2.13) which we can use as a starting point for finding this common denominator. It separates nine processes: 
\begin{enumerate}
    \item Business/use case modelling: Modelling how a system may be used
    \item Requirements: Actors and use cases are considered to model requirements
    \item Analysis and design: The architecture and functionality is modelled and documented
    \item Implementation: The design is realized with code
    \item Testing: Fulfillment of requirements is verified
    \item Deployment: The system is packaged and distributed to users
    \item Configuration \& change management: Changes to the system are tracked and managed
    \item Project management: The overall development process is managed
    \item Environment: Tooling for developers is distributed
\end{enumerate}

For the purpose of creating extensions for explorable authoring tools, we can disregard the the use case modelling, requirements, and project management processes here as they are independent of any interface provided by a tool. We also view configuration \& change management as external to the authoring, as for projects such as extensions, this is commonly achieved with a version control system such as Git. Furthermore, we don't consider tooling as its own requirement, but rather as a part of supporting the implementation. Finally, we extend the design process to include documentation.
\begin{enumerate}[leftmargin=2.25cm, label=EAR-DF\arabic*]
    \item \textit{Design/Document}: Model the architecture and functionality of an extension
    \item \textit{Implement}: Realize the functionality of an extension with code
    \item \textit{Debug/Test}: Find and fix issues with an extension manually or automatically
    \item \textit{Deploy}: Publish, update, deprecate and remove extensions
\end{enumerate}

\subsection{Authoring \& Development Experience}
Evaluating the user experience (UX) of a system is a well-discussed research problem in the area of human-computer interaction. There are many different scales to measure usability to choose from (compare Fig. \ref{tab:usability_scales}) An early, well established example of an instrument for measuring usability is the 10 item System Usability Scale (SUS). Commonly known as a "quick and dirty" way to measure usability, it returns a single overall result. While the SUS is still commonly used in some settings, the lack of sub scales limits the insights to be gained from evaluating. For example, the single value result does not allow researchers to differentiate between the correlated, but different factors of usability and learnability. There are several even more quick scales such as the ASQ, NPS, SEQ, and SMEQ - these are more applicable to market research settings where users may only be willing to answer a handful of questions.  Of course, these quick instruments can't offer any sub scales.

In a educational technology research setting, a more complex questionnaire with sub scales is viable. One such well-established scale is the 26 item User Experience Questionnaire (UEQ). The UEQ is available in more than 30 languages and measures usability across 6 sub scales (Laugwitz et al. 2008): Attractiveness, perspicuity, efficiency, dependability, stimulation, and novelty. For more specific applications, a modular framework to construct questionnaires based on the UEQ is available, named UEQ+ \cite{schrepp_measuring_2021}. This both extends the scales applicable to every system (with usefulness, adaptability, intuitive use, and trust), and introduces system-specific scales, for example for GUIs or voice assistants.

\begin{table}[H]
    \centering
    \begin{tabular}{llll}
         \textbf{Scale}&  \textbf{First published in}&  \textbf{Items}& \textbf{Subscales}\\
         After Scenario Questionn. (ASQ)&  \cite{lewis_ibm_1995}&  3& 0\\
 System Usability Scale (SUS)& \cite{brooke_sus_1996}& 10&0\\
 Post-Study System Usability Questionn. (PSSUQ)& \cite{lewis_psychometric_2002}& 19&3\\
         Net Promoter Score (NPS)&  \cite{reichheld_one_2003}&  1& 0\\
 User Experience Questionn. (UEQ)& \cite{holzinger_construction_2008}& 26&6\\
         Single Ease Question (SEQ)&  \cite{sauro_comparison_2009}&  1& 0\\
         Subjective Mental Effort Questionn. (SMEQ)&  \cite{sauro_comparison_2009}&  1& 0\\
 Visual Aesthetics of Websites Inventory (VisAWI)& \cite{moshagen_facets_2010}& 18&4\\
    \end{tabular}
    \caption{Usability Scales (adapted from \cite{hinderks_developing_2019})}
    \label{tab:usability_scales}
\end{table}

On the other hand, a consensus in evaluating the developer experience (DX) has yet to emerge. Lee \& Pan \cite{lee_evaluation_2021} outline that DX is in fact a special case of UX. This insight explains the large overlap between models of DX and models of UX. Consider their questionnaire based on the model of Fagerholm \& Münch \cite{fagerholm_developer_2012} and several studies from the area of acceptance research: In the questions listed, we can find almost every aspect of the UEQ+. For example, one aspect of the UEQ+ is efficiency, which appears in Lee \& Pan's items 21 ("it improves the efficiency") and 15 ("the platform is fast").

Given this research into UX and DX, we can make use of the idea that developers can also be seen as users. Developers of an extension for an authoring tool are, in other words, users of a specific subset of functionality the tool provides. Namely, they use the documented interfaces and parts of the tool itself to complete their task of developing an extension. For example, while perspicuity (ease of learning) in the context of the authoring experience may be achieved by a clearly designed UI with helpful tooltips, in the context of the development experience, it may mean adding type hints and documentation for an API.  Making use of this perspective and considering the UEQ+ model, we can divide our users into authors and developers and formulate parallel authoring and development experience requirements:
\begin{enumerate}[leftmargin=2.75cm, label=EAR-AX/DX\arabic*]
    \item \textbf{Efficiency}: Develop/author without unnecessary effort
    \item \textbf{Usefulness}: Fitness of purpose of the tool for authoring/development
    \item \textbf{Perspicuity}: Ease of learning to develop/author with the tool
    \item \textbf{Adaptability}: Customizing the tool for authoring/development
    \item \textbf{Dependability}: Having control of the tool for authoring/development
    \item \textbf{Intuitive use}: Immediate usefulness of the authoring tool for tasks
    \item \textbf{Stimulation}: Fun while using the tool for authoring/development
    \item \textbf{Novelty}: Creativeness of the design of the tool for authoring/development
    \item \textbf{Trust}: Safekeeping of documents/code when authoring/developing
    \item \textbf{Attractiveness}: Overall impression of authoring/development
\end{enumerate}

\section{Future Work: Operationalization of the EAR model} \label{futurework}
The previous section provided a catalogue of requirements called the EAR model. This sections adds a preliminary discussion on how the EAR model can be operationalized, that is, how to measure the degree to which each of the requirements is fulfilled. It is meant as an extended `related work` section, and further work is needed to develop this research design.

\subsection{Step 1: Evaluate authoring \& development functionality with qualitative analysis}
For the functionality requirements, we can view AF5-AF12 as a two-dimensional matrix (Element Type $\times$ Functionality), as there are individual challenges in supporting each functionality for each element type. This matrix is supplemented with fields for AF1-AF4, which apply to the whole document. Finally, the matrix also includes fields for the development functionality (DF1-DF4). Consider for example H5P, the main state-of-the-art system for authoring explorables (see Tab. \ref{tab:ear_h5p} as a speculative example). Systematically checking for each function and element type, we can tell that certain element types are well supported (such as text and headings), while many element types only enjoy limited support (rich text, tables, images, etc.) and yet others are not supported at all (disclosures, formulas, and graphics). This check could be performed as a qualitative analysis \cite{mayring_qualitative_2014} by one or more annotators using the same scale (such as the four-valued nominal scale of `no support', `code-only support', `limited support', and `full support' used in Tab. \ref{tab:ear_h5p}), using multiple annotators and calculating inter-annotator agreement for better quality.
\begin{table}[H]
    \centering
    \captionsetup{justification=centering}
    \begin{tabular}{r>{\centering\arraybackslash}p{0.07\linewidth}>{\centering\arraybackslash}p{0.07\linewidth}>{\centering\arraybackslash}p{0.07\linewidth}>{\centering\arraybackslash}p{0.07\linewidth}>{\centering\arraybackslash}p{0.07\linewidth}>{\raggedright\arraybackslash}p{0.07\linewidth}>{\centering\arraybackslash}p{0.07\linewidth}>{\centering\arraybackslash}p{0.07\linewidth}}
         \underline{Content}&  \textbf{Add/
rem.}&  \textbf{Edit}&  \textbf{Arrange}&  \textbf{Copy/
paste}&\textbf{Style} &\textbf{Gener.}&  \textbf{Attrib.}& \textbf{Licen.}\\
         Metadata&  \faCheck&  \faCheck&  \faCheck&  \faCheck& &&  &  \\
         Plain Text&  \faCheck&  \faCheck&  \faCheck&  \faCheck& &&  \faAsterisk&  \faAsterisk\\
         Rich Text&  \faAsterisk&  \faAsterisk&  \faAsterisk&  \faAsterisk& &&  \faAsterisk&  \faAsterisk\\
         Headings&  \faCheck&  \faCheck&  \faCheck&  \faCheck& &&  \faAsterisk&  \faAsterisk\\
         Sections&  &  &  &   & &&  &  \\
         Lists&  \faCheck&  \faCheck&  \faCheck&  \faCheck& &&  \faAsterisk&  \faAsterisk\\
         Tables&  \faCheck&  \faCheck&  \faAsterisk&  \faAsterisk& &&  \faAsterisk&  \faAsterisk\\
         Forms&  \faCheck&  \faCheck&  \faAsterisk&  \faAsterisk& &&  \faCheck&  \faCheck\\
         Disclosures&  &  &  &   & &&  &  \\
 Formulas& & & &  & && &\\
 Graphics& & & &  & && &\\
 Images& \faCheck& \faAsterisk& \faAsterisk& \faAsterisk& && \faCheck&\faCheck\\
 Audio& \faCheck& & \faAsterisk& \faAsterisk& && \faCheck&\faCheck\\
 Videos& \faCheck& & \faAsterisk& \faAsterisk& && \faCheck&\faCheck\\
 Documents& \faCheck& & \faAsterisk& \faAsterisk& && \faCheck&\faCheck\\
 Scripts& & & &  & && &\\
 Themes& & & & & & & &\\
 Widgets& \faCheck& \faAsterisk& \faAsterisk& \faAsterisk& && \faCheck&\faCheck\\
 & & & & &  && &\\
 \underline{Document}& \textbf{Create}&\textbf{Save/l./s.}& \textbf{Undo/r.}&\textbf{Extend}& &&&\\
 & \faCheck& \faCheck& \faCheck& \faAsterisk&  && &\\
 & & & & & & & &\\
 \underline{Extension}& \textbf{Design/d.}& \textbf{Impl.}& \textbf{Debug/t.}& \textbf{Deploy}& & & &\\
 & \faAsterisk& \faCode& \faCode& \faCheck& & & &\\
    \end{tabular}
    \caption{Speculative EAR Functionality Matrix of H5P/Lumi (v1.26, Dec. 2023)\\\_=No support, \faCode=Code-only support, \faAsterisk=Limited Support, \faCheck=Full support}
    \label{tab:ear_h5p}
\end{table}

\subsection{Step 2: Evaluate the development experience with a developer study}
To evaluate the development experience, we need to first of all consider that the development of an extension happens in a much longer time frame than authoring. While a useful document may be authored in a few hours, the development of an extension can span weeks or months (or even years when considering maintenance). Also, we expect the population of possible developers to be smaller than the population of possible authors, since developers need programming skills. For these reasons, we need to choose a different approach than with investigating the authoring experience.
One approach is to perform expert interviews \cite{bogner_interviewing_2009} with developers of open source extensions after release. This could provide insight into the whole process from start to finish. The interview can be prefaced by having the interviewee fill out an UEQ+ focused on the experience with the extension interface. The following interview can be separated into discussing each development functionalities (DF1-DF4), probing for the usability of each functionality with specific questions. The interview data can be supplemented with the source code repository for explication.

\subsection{Step 3: Evaluate the authoring experience with a user study}
Unlike with the development experience, with authoring, the whole authoring process can be observed in an experiment. To design a user study, a researcher should create a set of authoring tasks to complete. For example, a task could be to create a quiz where the learner needs to choose the right image for a given piece of foreign language vocabulary. Authoring this quiz would requires the adding, editing and arranging of text (possibly also rich text), images, and forms. The set of tasks created should ideally cover the whole functionality matrix to provide insight into the overall usability. In a user study, participants could then complete each task and then receive a UEQ+ questionnaire to fill out. Additionally, researchers could use screen recordings and the documents produced during the tasks as supplementary qualitative data to explain the results concerning usability. Methods such as think-aloud protocols \cite{krahmer_thinking_2004} could also be employed for smaller numbers of participants.

\section{Conclusion} \label{conclusion}
In this paper, we introduced a set of requirements for explorable authoring tools. The interactive, multimodal and open nature of explorables makes them both a promising medium for education and a difficult one to author, requiring web development expertise, excluding most educators and students from the authoring and remixing process. To evaluate current and future tools for authoring explorables, we, on the basis current models in usability and engineering, empirical studies and domain modelling, derived the Explorable Authoring Requirements (EAR). We finally outlined a future operationalization of EAR.

While this work is a useful step towards evaluating and comparing explorable authoring tools, it has some limitations. First, the proposed method of evaluating the authoring and development functionality is not standardized, leaving it unexamined in terms of scientific quality criteria such as validity, objectivity and reliability (although qualitative analysis with inter-annotator agreement is a well-accepted practice). Second and similarly, the application of the UEQ+ to the development experience, while theoretically sound, is not tested. Third, this study does not offer a full study that measures the EAR for a given tool, only the outline of an operationalization.

Future work could remedy most of these listed shortcomings by applying the EAR model to different tools, as outlined in the previous section. We intend to elaborate on the outlined plan and investigate the functionality and experience for authoring and development with the WebWriter tool in separate studies.  In general, studies applying this model may yield useful data sets that may vary depending on the target group participating in the user study, the tasks chosen by the researchers, and so on. With different data sets, it also becomes possible to properly compare tools and to spot room for innovation where all tools currently available are lacking.

\printbibliography

\end{document}